# Sustainable and effective antimicrobial surface based on cellulose thin films

Shaojun Qi[1], Ioannis Kiratzis[1], Pavan Adoni[1], Zania Stamataki[2], Aneesa Nabi[1], David Waugh[3], Javier Rodriguez Rodriguez[1], Stuart Clarke[4], Peter J Fryer[1], and Zhenyu J Zhang[1*]

[1] School of Chemical Engineering, University of Birmingham, Birmingham B15 2TT, UK

[2] Institute for Immunology and Immunotherapy, University of Birmingham, Birmingham B15 2TT, UK

[3] School of Mechanical, Aerospace and Automotive Engineering, Coventry University, Coventry CV1 2JH, UK

[4] Yusuf Hamied Department of Chemistry, Cambridge University, Cambridge CB2 1EW, UK

[*] Corresponding author. E-mail Z.J.Zhang@bham.ac.uk

## Abstract

Despite well-established surface cleaning and disinfection solutions, it remains challenging to provide long lasting protection to surfaces whilst minimising the use of harmful substances, as seen in conventional anti-microbial products which pose threats to the urban environment and biodiversity. In the present work, we developed a sustainable and effective antimicrobial surface film based on Micro-Fibrillated Cellulose. The resulting porous cellulose thin film is barely noticeable to human eyes due to its sub-micron thickness, of which the coverage, porosity and microstructure can be modulated by the formulations developed. Using goniometers and a quartz crystal microbalance (QCM), we observed a threefold reduction in water contact angles and accelerated (more than 50% faster) water evaporation kinetics on the cellulose film. The thin film exhibits not only a rapid inactivation effect against SARS-CoV-2 in 5 minutes, following deposition of the virus loaded droplets, but also an exceptional ability to reduce contact transfer of liquid, e.g. respiratory droplets, onto surfaces such as artificial skin by more than 90%. It also exhibits excellent antimicrobial performance in inhibiting the growth of both gram-negative and gram-positive bacteria (*E.coli* and *S.epidermidis*) due to the excellent porosity and hydrophilicity. Additionally, the cellulose film shows nearly 100% resistance to skin scraping in dry condition thanks to its strong attachment to the substrate, whilst good removability once wetted, suggesting its practical suitability for daily use. Importantly, the coating can be formed on solid substrates readily by spraying and requires solely a simple formulation of a plant-based cellulose material with no additives, rendering it a scalable, affordable and green solution for antimicrobial surfaces. Implementing such cellulose films could thus play a significant role in controlling future pan- and epidemics, in particularly during the first phase when appropriate medication needs to be developed.

## Introduction

A variety of infectious pathogens, including respiratory viruses and disease-causing bacteria, can spread via surface fomites [1]. The transmission of SARS-CoV-2, for example, can take place by contact with fomites that were contaminated by infectious respiratory secretions or droplet nuclei [2]. In a recent study by Public Health England, it is reported that viable virus persisted for long periods of time on hydrophobic surfaces, eg. surgical masks and stainless steel up to 7 days [1]. It has been shown that a contact of only 5 seconds is sufficient to transfer 31.6% of the influenza virion load to the hands [3]. A recent study also shows that about 15% of SARS-CoV-2 was picked up by the finger though a light touch of the fomite [4]. Surface disinfection is a well-established practice to intercept transmission of fomite-induced diseases. Chemical disinfectants, such as conventional bleach with appropriate dilution [5], ethanol, and treatment with dried hydrogen peroxide [6] are commonly used to sanitise surfaces with an effect in minutes [6-8]. However, many disinfectants contain harmful and corrosive substances such as chlorine bleach, phenolics and quaternary ammonium compounds [9-10]. For example, chlorine disinfectants irritate the mucous membranes of the respiratory and digestive systems [11] and can cause acute toxicity or even death to both terrestrial and aquatic wildlife [12]. Unfortunately, the prevailing practice of large-scale, frequent, indiscriminate and sometimes excessive application of disinfectants amid COVID-19 is worrying, posing threats to the urban environment, biodiversity and the public health [10]. The frequency (69.3%) and amount (74.2%) of cleaning product usage both increased significantly since the pandemic, according to a survey in autumn 2020 in Turkey [13]. Data from the CDC in the United States also shows 20.4% more reports of human poisoning due to exposures to cleaners and disinfectants following unsafe use in the first three months of the COVID-19 outbreak in January 2020 [14]. Whilst physical methods such as UV irradiation [15] and heat treatment [7] are effective in disinfecting surfaces, they are less practical for use on a daily basis and at large scale. Self-disinfecting surface coatings incorporating selected metal elements, notably copper and silver, exhibit virucidal properties by proactively disrupting the disulphide bonds of virus proteins [16] and/or releasing reactive oxygen species (ROS) that damage the nucleic acids of the virus [4, 17-19]. However, they are limited by either the high cost or the complicated procedures that are too often inaccessible for the public, not to mention the involvement of multiple metal salts and irritating chemicals that may sacrifice the sustainability. Therefore, it is important to develop technologies and hygiene products that can effectively, continuously, yet sustainably, function against notable pathogens, with little impact on the environment, which is an unmet characteristic of most of the current designs and products. It was estimated that the market for sustainable cleaning products could be 110 billion U.S. dollars per annum in 2025 [20].

It is also crucial to note that the conventional chemical disinfectants and antiviral surface designs primarily target the enveloped virus, either its membrane or nucleic acids, rendering them non-functional. However, it is the respiratory liquid that makes first contact with the recipient surface [21]. There has been literature reporting that coronaviruses such as SARS-CoV-2 can remain viable for weeks in water and moist environments [22-23]. Previous studies have shown a clear correlation between the evaporation kinetics of respiratory droplets on a surface and the virus persistence time [24]. The underpinning principle of virus inactivation developed here is thus to target the respiratory droplets as opposed to the virions within. Porous surfaces enable a liquid imbibition process which is much faster than diffusion-limited evaporation and spreads and drains the droplet volume quickly [24]. The virus is dried into the interior of the porosity [25] and loses infectivity by up to 100-fold during drying [26]. With the virus being trapped and adhering to the pores below surface, microbial transfer through finger touching or rubbing can also be significantly reduced [4]. Such design principles for antimicrobial coatings, without involvement of chemical additives, are attractive as a passive surface hygiene approach to inhibit the transmission of infectious diseases. The limiting factor, nevertheless, is the lack of manufacturing capability to scale up and implement such coating in a sustainable and cost-effective manner.

Inspired by the clinical evidence and previous work on porous surface coating, we have developed cellulose-based surface films that are hydrophilic and porous, so that respiratory droplets could be absorbed immediately. The evaporation kinetics of water and a simulated respiratory fluid on the cellulose coated surfaces were investigated. The efficiency of the films in reducing droplet contact transfer and the mechanical stability of the films were evaluated. Antiviral efficiency was tested by deploying an infectious SARS-CoV-2 culture. The film developed could serve as a sub-micron porous coating that offers multiple functions, e.g. antiviral, antimicrobial, and active carriers, so that the film could be used for regular contact surfaces.

## Results

### Morphological characteristics of Micro-Fibrillated Cellulose film

Porous films were fabricated on glass substrates by either spin coating or spraying aqueous suspension of Micro-Fibrillated Cellulose (MFC). A dry film was formed on the substrate within 10-30 seconds after applying the suspension. It is suggested that hydrogen bonds naturally form between the fibrils which can immobilise the network and enable sufficient adhesion to the substrate to make them durable. Figure 1 presents the morphologies of two representative thin films with fine details. We observed a porous and interconnected web-like structure, consisting of individual cellulose fibrils and a small fraction of pulp bundles are observed. The two coating methods produced a coverage of 91.2% (spin coating, referred to as MFC-I) and 43.9% (spraying, MFC-II) respectively on the substrate.

Based on the images acquired by atomic force microscopy, the height distribution histogram was compiled (Figure 1f). The prepared MFC-I and MFC-II films were 1.2 µm and 300 nm thick, respectively, rendering them barely noticeable to human eyes (Figure 1g). Surface roughness, waviness, and porosity levels were quantified and are compared in Table 1. MFC-I possesses higher surface roughness, but a lower mean pore size compared with MFC-II. It is worth noting that the morphological characteristics of the cellulose films in this work could be effectively modulated by varying the fabrication conditions. For example, an increased speed for the spin coating process reduced the roughness and increased the porosity of the resulting film, whilst more drops of cellulose suspension led to increased surface roughness and reduced porosity levels (Figure S1, Supplementary Materials).

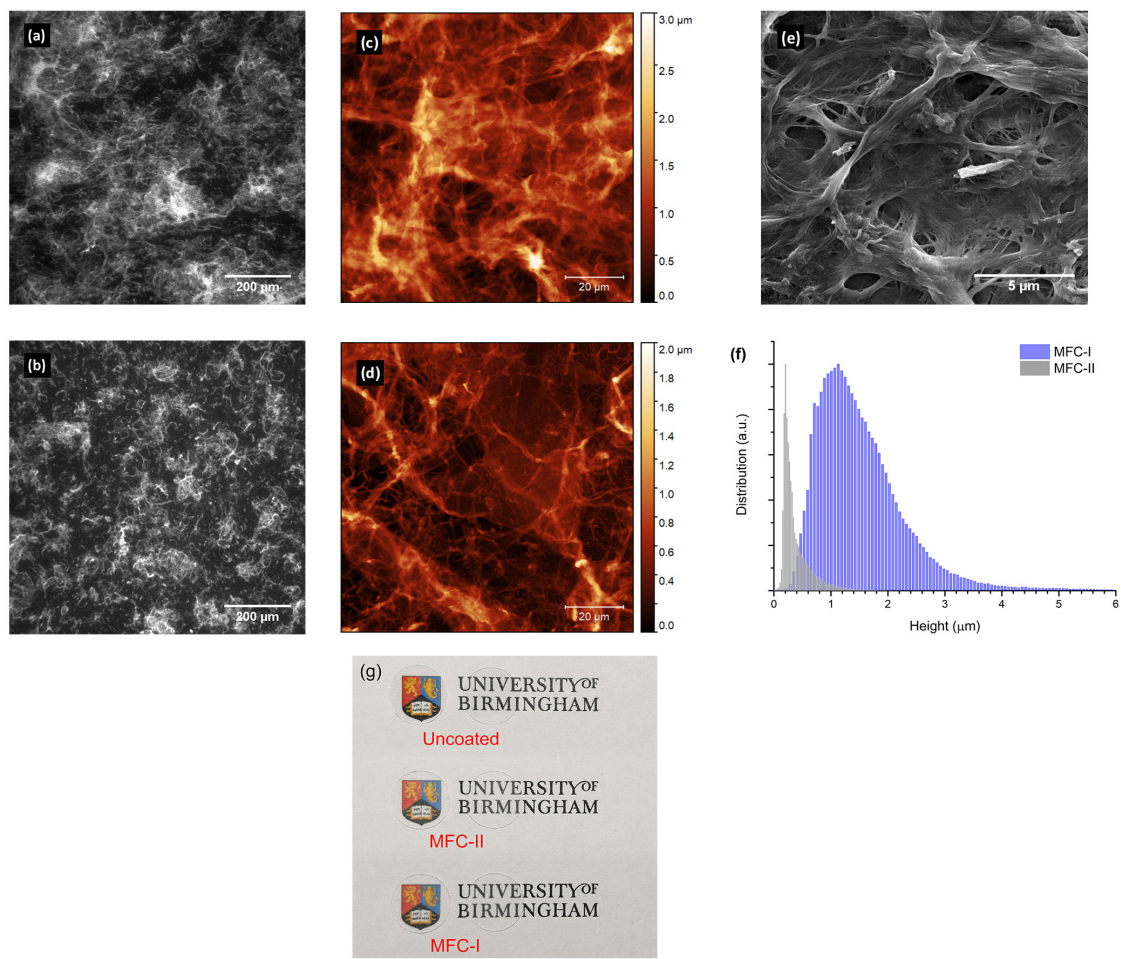

Figure 1. Microscopic characterisation. a-d, Morphology of the cellulose thin films observed by an (a and b) optical microscope and (c and d) AFM. The sample in (a) and (c) was prepared by spin coating (i.e., MFC-I); the sample in (b) and (d) was by spraying (MFC-II). e, SEM showing a representative view of the porosity (scale bare 5 μm). f, Height histograms of the two cellulose films. g, A photo of uncoated glass and those coated with the cellulose thin films demonstrating their transparency against natural light.

Table 1. Surface parameters of the MFC thin films

|  | MFC-I | MFC-II |
| --- | --- | --- |
| Roughness $R_a$ (nm) | 247 ± 54 | 97 ± 10 |
| Waviness $W_a$ (nm) | 460 ± 135 | 144 ± 20 |
| Porosity (% of projected area) | 31.67 ± 6.51 | 58.06 ± 9.16 |
| Mean pore size (μm) | 5.82 ± 0.53 | 10.06 ± 2.30 |

Evaporation of droplets

Surfaces bearing virus-laden droplets can serve as media for the spread of the virus. Previous studies suggest that porous surfaces hold advantages in disrupting the surface viability of virus due to much faster droplet evaporation, facilitated by a capillary imbibition effect, as opposed to the slow diffusion-limited evaporation on impermeable surfaces [24]. As such, creation of porous and fluid-absorbing coatings is an advantageous strategy for antiviral surfaces and devices. They offer unique practical and environmental benefits missing in other antiviral solutions, such as metallic, polymer and multi-layered nanoparticle coatings [27].

Figure 2a compares the initial water contact angles of two distinct droplet sizes (60 μm and 1 mm in diameter) on the two cellulose thin films, taking a flat glass surface (uncoated, only cleaned with ethanol prior to testing) as the control. The 1 mm droplets exhibited contact angles similar to the 60 μm ones on each type of substrate. Against droplets of the two length scales, MFC-I and MFC-II showed much smaller contact angles which are about 1/3 and 1/2 of that on uncoated glass, respectively. Apart from the difference in material composition, this reduction in contact angles could be attributed to the hydrophilic and thus more wettable surface of the cellulose architecture, which is consistent with the mechanism suggested by previous study that the surface porosity could introduce an imbibition effect to distribute the water through the pores due to capillary force, which then evaporates at a fast rate [24].

Rapid droplet spreading and imbibition on the cellulose thin films was observed using a contact angle goniometer (Figure 2b). On a flat, solid glass surface allowing solely evaporation, a droplet of 60 μm diameter remained discernible for 15s. That time was halved on both cellulose films. More significantly, the duration for which 1 mm diameter droplets were present above the surface was reduced from 498 s on flat glass to 110 s when cellulose films were in place. Note that as water is wicked down the surface of the cellulose films some will be absorbed within the porous structure. The results in Figure 2b suggest that the cellulose thin films can effectively accelerate the loss of water by wicking of the aqueous droplets, hence also reducing the time window during which indirect contact transfer could take place.

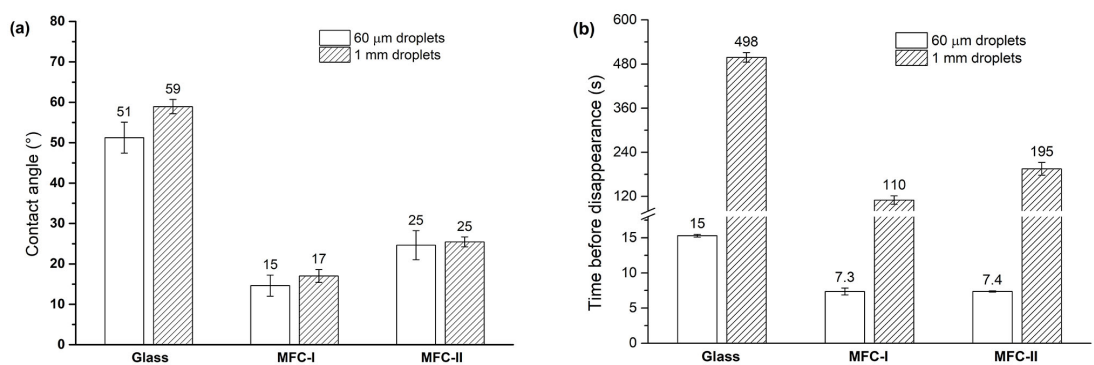

Figure 2. **Contact angle measurements**. a, Initial contact angles of water droplets of two different sizes against glass and the MFC thin films. b, Time before droplets disappeared from the surface, determined as the point when the droplet is no longer discernible by the side view camera of the contact angle goniometer.

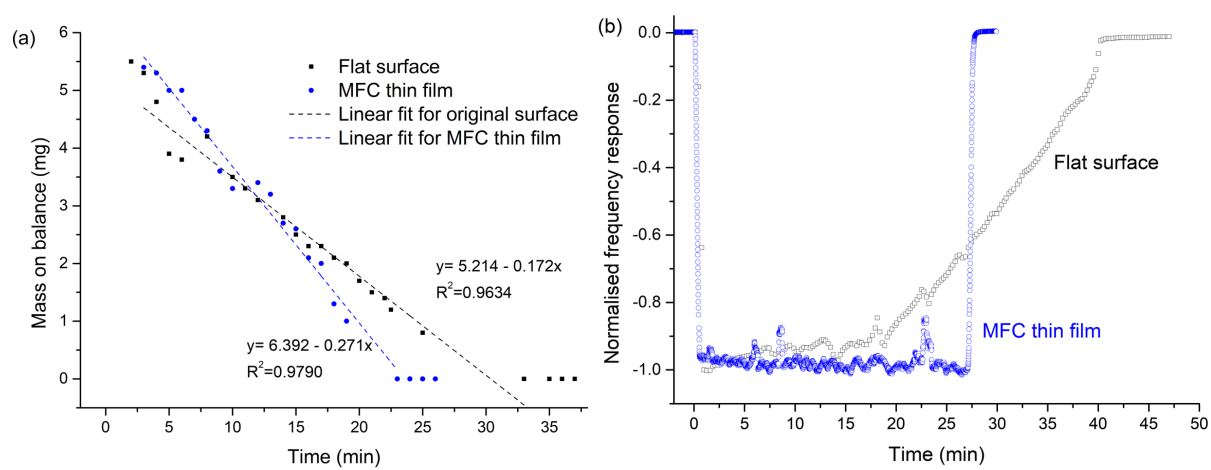

Figure 3. **Laboratory mass balance (a) and QCM (b) analysis of droplet evaporation**. Water droplets of 5 were placed on the mass balance and the quartz crystal surface and left evaporating until full dryness.

The effect of the porous cellulose film on water evaporation kinetics was demonstrated on a laboratory balance with 0.1 mg readability. All experiments were conducted at room temperature and 48-50% relative humidity. Figure 3a plots the weight reading history of a water drop (initial volume 5 µL) as a function of time. The sessile water drop evaporated clearly faster (0.271 vs. 0.172 mg·min$^{-1}$) on cellulose coated surface than on flat, uncoated surface. The linearity of the mass loss on flat, uncoated surface (black squares and fitted line) indicates an evaporation process governed by diffusion, with the highest evaporation flux occurs at the contact line [24, 28]. The evaporation rate of sessile droplets thus shall be proportional to the contact radius [29]. The shifted evaporation kinetics observed on the cellulose film (blue dots and fitted line) reflects the effect of the surface properties. In the present work, the evaporation of the sessile droplets was simultaneously monitored using a quartz crystal microbalance (Figure 3b), which is highly sensitive to changes in the temporal droplet radius [30-31]. We observed that the droplet evaporating on the flat surface initially fell in a constant contact radius (CCR) regime, during which the liquid-solid contact line is pinned whilst the contact angle and the droplet height decrease as the evaporation continues, followed by a depinning period where the contact line contracted until the completion of the evaporation. In contrast, for the surface covered by the cellulose thin film with hydrophilic and porous properties, the droplet remained evaporating in a constant contact radius mode for most of its lifetime due to the capillary-driven spreading and imbibition in the presence of the porous medium, leading to a much accelerated evaporation process [28].

**Anti-microbial testing**

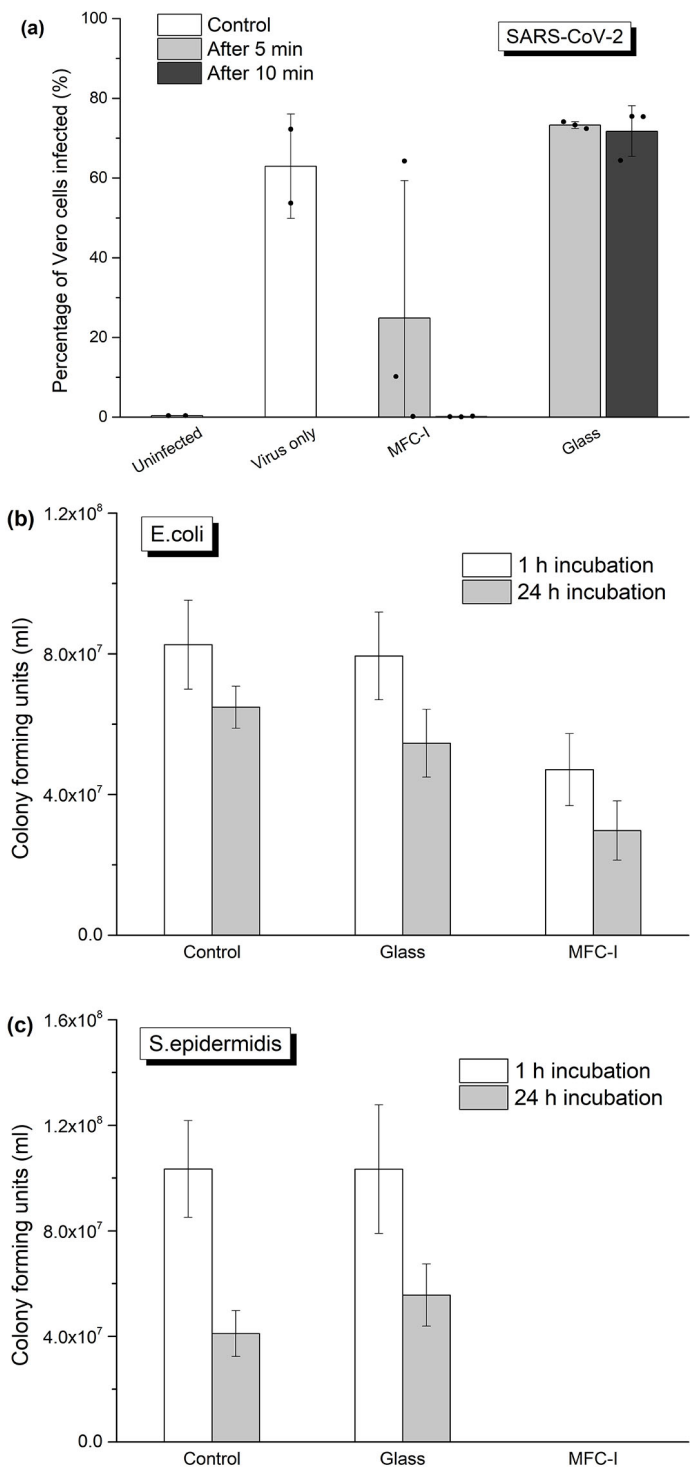

Figure 4. **Anti-microbial testing results.** a, In vitro inactivation testing of SARS-CoV-2 on cellulose film and glass slide. Drops of the virus culture medium were left on the porous cellulose film (MFC-I) and planar laboratory grade glass coverslips. Any infectious virions were recovered from these surfaces either 5 or 10 min after droplet deposition and transferred to target cells for infection. A control group (i.e. virus medium added to the cell culture directly without pre-exposure on any substrates) is also shown. b and c, Viability of two bacteria (*E.coli* and *S.epidermidis*) after incubation on the cellulose thin film and glass for 1 h and 24 h.

The cellulose films' ability to inhibit surface transmission of SARS-CoV-2 was studied through in vitro infection of Vero cells [32]. To replicate the transmission process of surface fomites, virus containing droplets of 1 mm diameter were deposited on the surfaces and left for 5 or 10 min before the recovery and infection procedures to determine the amount of viable virus. Figure 4a shows the viability of SARS-CoV-2 as a function of substrate characteristics and droplet residence time. The direct contact of the virus medium with Vero cells ('virus only') led to an infection rate of up to 70%. A threefold reduction of infected cells was achieved when droplets of the virus medium were left on the cellulose thin film (MFC-I) for 5 min. After a residence time of 10 min, the infectivity of the virus-containing droplets was minimal and comparable to the uninfected group, which indicates a major loss of viability for the virus. In contrast, the viability of SARS-CoV-2 was unaffected if the virus-containing droplets were deposited on a plain glass surface even after 10-min residence. This comparison suggests a clear and quick virus inactivation effect of the cellulose film in the present study. We interpret this virus inactivation mechanism primarily as a consequence of the accelerated water evaporation on the cellulose films. The virus-containing droplets of 1 mm diameter would have been dried out completely after 5 min on the cellulose coated substrate, as evidenced in Figure 3a and Figure S2a, leaving the virions exposed to the ambient environment and prone to disruption. The virus drops could remain hydrated, however, after 10 min on a flat, non-porous surface (Figure 3a, Figure S2a).

The effectiveness of the cellulose film in inactivating bacteria was assessed using two representative pathogens: *E.coli* (Figure 4b) and *S.epidermidis* (Figure 4c), alongside two sets of benchmarks: control (bacteria incubating in a broth) and those incubating on glass slides. Whilst the bare glass surface showed statistically no effect on the viability of both bacteria incubating on its top when compared to the control group, the viability of *E.coli* was reduced by 43% and 54% when incubating on the cellulose thin film for 1 and 24 h, respectively. Meanwhile, we found that S. epidermidis was especially vulnerable to the cellulose coated surface, with complete loss of viability after incubation for 1h. The antibacterial effect of the cellulose film can be likely attributed to its hydrophilicity and porosity. It is known that hydrophilic surfaces can prevent the attachment of bacteria and thus inhibit the formation of a biofilm due to the presence of a water molecule layer which disfavours the adsorption of bacteria. The cellulose film in this work enables the generation of a water barrier layer immediately and uniformly upon the deposition of the bacteria culture, which gives rise to the observed antibacterial effect.

The results in Figure 4 imply that instead of modulating merely the hydrophobicity or hydrophilicity of a surface, creating hydrophilic yet porous surfaces, as demonstrated in the present work, could be a more effective strategy for the inactivation of respiratory viruses as well as disease-causing bacteria. Furthermore, the high surface area and rich hydroxyl sites that are naturally present on the cellulose fibrils can facilitate many add-on antimicrobial solutions through surface functionalisation.

**Contact transfer and mechanical stability**

A recent study has reported that SARS-CoV-2 can be transferred to an artificial finger through a brief, light contact with contaminated surfaces and the transfer of virus was found to be much less from porous than from non-porous substrates. The authors hypothesise that porosity plays the key role by allowing penetration and thus less transfer of respiratory liquid, which in turn reduces the transfer of virus [4]. Despite this plausible hypothesis which is supported by the virus titer for the finger, there is no study yet in the literature visualising the effect of porosity on contact transfer of respiratory droplets. In the present study, we performed a series of contact measurements between an artificial skin and substrates with or without our cellulose films, in the presence of respiratory liquid (an artificial saliva), of which fluorescence micrographs are presented in Figures 5.

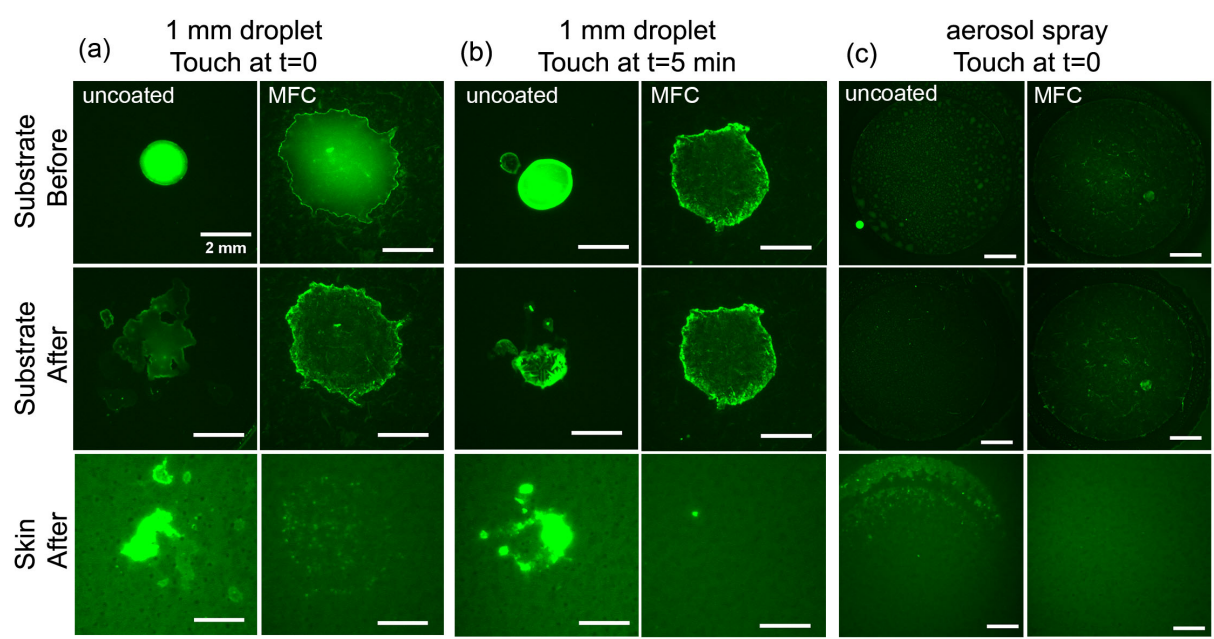

Figure 5. Contact transfer tests. Artificial saliva droplets (a and b) and aerosol (c) were deposited on the substrates either with or without our cellulose films (MFC) and compression contact with an artificial skin was made at a set time. a, Contact made immediately after the deposition of a 1 mm droplet. b, Contact made 5 min after the deposition of a 1 mm droplet. c, Aerosol of the artificial saliva was sprayed onto the substrates and contact was made immediately following the aerosol spray. Scale bars 2 mm in all images.

Figure 5a shows that upon deposition (top line in the plot) of a 1 mm diameter artificial saliva droplet on the host substrates, the droplet spread 400% wider on the cellulose coated surface compared with that on the uncoated glass surface. The contact was then made immediately following the droplet deposition. After the contact (middle and bottom lines in Figure 5a), an intense and a large area of fluorescence was observed on the artificial skin; this must have left the uncoated substrate, suggesting a considerable amount of transferred liquid. In contrast the artificial skin after contacting the cellulose coated surface showed only faint fluorescence stains, of which the intensity is significantly weaker, and the contaminated area is approximately 90% less. Figure 5b shows the scenario where a droplet residence time of 5 min was allowed before the contact was made. Prior to the contact, the droplet on the glass substrate was still hydrated whilst the droplet on the cellulose coated substrate was fully dry already, based on their morphologies as shown in the micrographs. Consequently, the artificial skin in contact with the uncoated glass was once again heavily soiled. In contrast, we observed nearly zero fluorescence stains on the artificial skin that had contact with the cellulose film. Figure 5c shows the contact transfer performance of the same substrates that were subject to sprays of artificial saliva aerosol. Figure 5c and the quantitative analysis (Figure S2) suggests that the cellulose thin film is also effective in suppressing the contact transfer of respiratory aerosols.

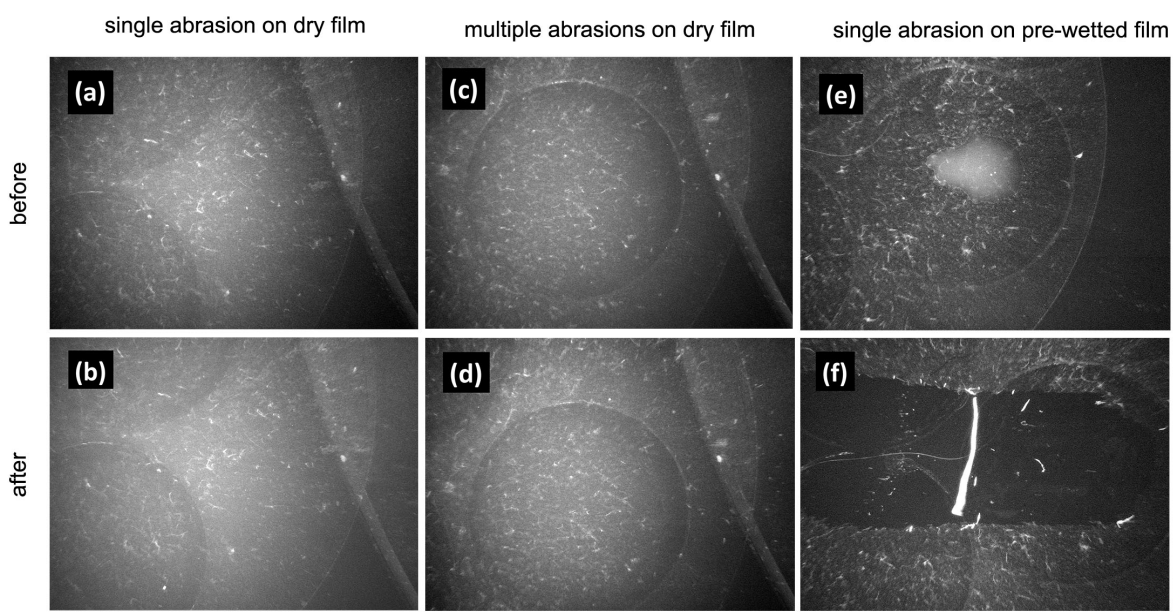

**Figure 5. Mechanical stability tests.** Morphology of the cellulose thin films before and after scraping tests under different wear conditions. a-b, single pass. c-d, reciprocating, multiple passes. e-f, single pass, against pre-wetted cellulose film. The testes were performed using an artificial skin covering the whole substrate surface.

The mechanical stability of the cellulose thin films was evaluated by means of scraping tests against an artificial skin under a normal compression load (The actual load history is shown in Figure S4). Images of the exact locations prior and post the abrasion tests were acquired by a bright field optical microscope. As shown in Figure 5a-d, the cellulose thin films did not show any noticeable damage after the abrasion under dry conditions, even after multiple cycles of back and forth scraping. The satisfactory mechanical strength of the thin film is attributed to the hydrogen bonding between cellulose fibrils and the supporting substrate (glass in the present work), which sufficiently immobilises the network of cellulose fibrils, providing a considerable resistance to occasional abrasions while it is dry. This suggests the suitability of such coating for high traffic objects such as door handles, handrails that are regularly in contact with human skin. The cellulose thin film, however, can be easily removed with a single abrasion, with the presence of a small quantity (1 mm droplet in the present work) of liquid (Figure 5e-f). The results imply that the cellulose film can not only be easily applied on common surfaces and maintain its strength and functionality in dry conditions, but also exhibit great removability once wetted, making it convenient and suitable for daily cleaning and disinfection practice.

Conclusion

A sustainable and effective antimicrobial surface thin film based on Micro-Fibrillated Cellulose has been developed. The porous cellulose film, with a submicron thickness, is barely noticeable to human eyes but can effectively render the surface hydrophilic (a threefold reduction in water contact angles) and accelerate (more than 50% faster) the evaporation of respiratory droplets which is shown by QCM analysis. The film exhibits not only a rapid inactivation effect against SARS-CoV-2 in 5 minutes following the deposition of the virus-containing droplets, but also an exceptional ability to reduce contact transfer of liquid, e.g. respiratory droplets, onto surfaces such as artificial skin by more than 90% as shown by contact transfer tests. Additionally, the thin film is also effective in inhibiting the growth of both gram-negative and gram-positive bacteria (*E.coli* and *S.epidermidis*). Moreover, with strong attachment to the substrate the cellulose film can provide nearly 100% resistance to skin scraping in dry condition, whilst good removability once wetted, suggesting its practical suitability for daily use. Importantly, the thin film can be formed on solid substrates readily by spraying and requires solely a simple formulation of a plant-based cellulose material. The flexibilities in controlling formulation, cellulose surface chemistry, coating process offer a powerful platform to deliver appropriate functionalities. These notable advantages of the cellulose film offer a scalable, affordable and green solution to contain infectious diseases that spread via respiratory fluid.

Acknowledgement

We thank the financial support by the Engineering and Physical Science Research Council (EP/V029762/1), and FiberLean Technologies for kindly providing Micro-Fibrillated Cellulose samples.

## Methods

Materials

A Micro Fibrillated Cellulose (MFC) aqueous slurry (solid content 7wt%, Fiberlean Technologies Ltd. UK) was diluted with ethanol (v/v = 1:5) and homogenised (SHM1 Homogeniser, Stuart, UK) for 3 min before use. Polyethyleneimine (PEI; 181978), Phosphate buffered saline (PBS; P4417), mucin (Type I-S; M3895), Bovine Serum Albumin (BSA; A9647), and Tryptone (T9410) were purchased from Sigma-Aldrich. Fluorescence dye Alexa Fluor™ 488 C5 Maleimide was purchased from ThermoFisher Scientific.

An artificial saliva solution was prepared in compliance with international standard ASTM E2197 and formed of three types of proteins (i) high molecular weight proteins (e.g. Bovine Serum Albumin, BSA), (ii) low molecular weight peptides (e.g. Tryptone), and (iii) mucous material (e.g. Mucin). In a typical process, the three protein solutions are prepared separately by adding 0.5 g BSA, 0.5 g Tryptone and 0.04 g mucin into 10 mL PBS individually. Each of them is then passed through a 0.22 µm pore diameter membrane filter, aliquot, and stored at either 4 ± 2°C or -20 ± 2°C. To obtain 500 µL artificial saliva solution, add 25 µL of BSA, 100 µL of mucin, and 35 µL of Tryptone to 340 µL PBS and mix well (magnetic stirring, 30 min) before use. The concentration of mucin in the resulting artificial saliva is 0.8 mg·mL$^{-1}$.

Nutrient agar (OXOID Ltd.) contained (per litre of deionised water) 1 g 'Lab-Lemco' powder, 2 g yeast extract, 5 g peptone, 5 g sodium chloride, 15 g agar (pH 7.4). Luria-Bertani broth contained (per litre of deionised water) 10 g sodium chloride, 10 g tryptone and 5 g yeast extract. Phosphate buffered saline (OXOID Ltd.) contained (per litre of deionised water) 8 g sodium chloride, 0.2 g potassium chloride, 1.2 g of di-sodium hydrogen phosphate and 0.2 g potassium dihydrogen phosphate. All solutions were autoclaved for 20 minutes before use.

Thin film fabrication

Glass coverslips (ϕ 10 mm, thickness 0.16-0.19 mm, Fisher Scientific, UK) were cleaned with ethanol and then placed in an oxygen plasma chamber (HPT-100, Henniker Plasma) at an oxygen flow rate of 10 sccm for 5 min. Polyethyleneimine solution (1% w/v in H$_2$O) of 70 µL was placed on the cleaned substrate. The loaded substrate was spun at 600 rpm for 30s on a spin coater (SPIN150i, APT GmbH), then accelerated at 500 rpm/s to 4000 rpm and spun for 60 s. MFC thin films were fabricated on such pre-treated surfaces by two different approaches, namely spin coating and spray coating. In the case of spin coating, 400 µL MFC suspension was added dropwise onto the sample spinning at 6000 rpm. In the case of spray coating, a manual cosmetic atomiser was used to apply the MFC suspension onto the stationary sample. A total of 40 sprays were made to obtain a uniform coverage.

Surface characterisation

Surface morphology of the thin films was examined by an atomic force microscope (AFM, Multimode, Bruker) with a tapping mode cantilever (NCHR-20, Apex Probes Ltd.) and a scanning electron microscope (Philips XL-30 FEG ESEM). Surface parameters were extracted from the scans (over a range of 1.4x1.0 mm) by a white light interferometer (WLI). Porosity levels of the thin films were evaluated semi-quantitatively using the image processing program Gwyddion and the integrated Watershed algorithm [33]. Contact angle measurements were carried out using two distinct droplet sizes, i.e. 60 µm and 1 mm in diameter (100 pL and 0.5 µL in volume), for which an optical contact angle instrument equipped with a picolitre dosing system (PDDS, DataPhysics Instruments GmbH) and a generic contact angle goniometer (Ossila Ltd.) were employed, respectively.

Quartz Crystal Microbalance

The evaporation behaviour of water droplets was studied using silicon dioxide coated QCM sensors (5 MHz 14 mm Cr/Au/SiO2, Quartz Pro, Sweden). The surfaces of the sensors were pre-treated and MFC thin films fabricated as mentioned in the 'Thin film fabrication' section. Water droplets of 5 µL were placed onto the QCM sensor by a micropipette. The frequency and energy dissipation history through the quartz sensors during the droplet landing and evaporation events was simultaneously monitored and recorded by a quartz crystal microbalance (NEXT, openQCM, Italy).

Antiviral analysis

We attempted to replicate virus droplets as in a sneeze from an infected person whereby 0.5 µL drops of medium containing SARS-CoV-2 (England 2 stock 106 IUml−1 (kind gift from Christine Bruce, Public Health England)) were placed on the testing materials and left at room temperature for either 5 minutes (semi-dry) or 10 minutes (dry). The deposited drops were evident immediately in the porous materials. We then retrieved any remaining infectious virus from the treated surfaces using 50 µl of cell culture medium on top of the viral drops, which were transferred to target cells for infection. We measured infection in Vero cells at 48 hours, by scoring the percentage of spike-expressing cells.

The Vero cells were washed with PBS, dislodged with 0.25% Trypsin–EDTA (Sigma Life Sciences), and seeded into 96-well imaging plates (Greiner) at a density of $10^4$ cells per well in culture media (Dulbecco's modified Eagle medium (DMEM) containing 10% FBS, 1% penicillin and streptomycin, 1% l-glutamine and 1% non-essential amino acids). Cells were incubated for 24 h to allow time for adherence, and were fixed in ice-cold MeOH after infection. Cells were then washed in PBS and stained with rabbit anti-SARS-CoV-2 spike protein, subunit 1 (The Native Antigen Company), followed by Alexa Fluor 555-conjugated goat anti-rabbit IgG secondary antibody (Invitrogen, Thermofisher). Cell nuclei were visualized with Hoechst 33342 (Thermofisher). Cells were washed with PBS and then imaged and analyzed using a ThermoScientific CellInsight CX5 High-Content Screening (HCS) platform. Infected cells were scored by perinuclear fluorescence above a set threshold determined by positive (untreated) and negative (uninfected) controls.

Bacterial testing

Strains of *E. coli* and *S. epidermidis* were incubated respectively in 10 ml L-B broth overnight in a 37°C incubator with shaking at 150 rpm. Both species were pelleted and washed with 10 mL PBS solution twice and suspended in PBS to $OD_{600}$ 0.1 (*E. coli*: 8.5 x $10^7$ cells/ ml, *S. epidermidis*: 10.3 x $10^7$ cells/ ml). To the coated slides, 20 ul of the bacterial culture was added, slides were placed in 24 well plates which were sealed with parafilm and incubated at 30°C for either 1 h or 24 h. After the set amount of time, surviving bacteria were recovered from the slides as follows. (i) Slides were placed into 0.7 ml PBS in 15 ml falcon tubes respectively and vortexed for 30 s. (ii) Slides were further physically scraped with a spatula and the residue mixed into the respective 0.7 ml solution from (i). (iii) Samples were then sonicated for 3 x 1 min in a bath sonicator (GT Sonic, 40 Hz, 100 W). Serial dilutions were then performed and 10 ul of the final dilution was pipetted onto nutrient agar plates which were left to soak into the agar for 30 mins. Plates were then incubated at 37°C overnight, after which colonies were counted to determine the antibacterial effect of the coatings.

Contact transfer and durability testing

To evaluate the transfer of respiratory liquid from the substrate, a piece of artificial skin was pressed against the sample surface either immediately following the loading of artificial saliva droplets or after a residence time of 5 min. Both 1 mm diameter droplets and aerosols of artificial saliva were employed. For the case of aerosol, a nebuliser (Omron C28P) ejected aerosolised droplets (mass median aerodynamic diameter 3.0 µm) towards the sample for 30 seconds. The substrate, either uncoated or MFC coated, was attached to a glass slide using carbon adhesive discs. The artificial skin was fixed on one end of an instrumented arm (Forceboard, Industrial Dynamics, Sweden AB) which in each test was driven smoothly towards the sample until a contact force of 2 N was reached. The artificial skin was then retracted from the sample surface. Each touch cycle lasted approximately 5 seconds. Figure S1 shows the set-up. The touched sample surface and artificial skin counter surface were both examined under a fluorescence microscope (Leica Z16 APOA). The artificial saliva solution was stained by Alexa Fluor 488, which facilitates the observation and quantification of the mucin contaminations on each side. The areas of the fluorescent protein stains after the touch were measured via image processing using ImageJ.

Lateral scraping tests were performed to assess the mechanical stability of the thin films under both dry and wet conditions. A piece of artificial skin was fixed on the level of the force board and lowered onto the substrate until a normal force of 2 and 4 N was reached. Both one-pass and reciprocating multi-pass tests were performed in ambient environment. An additional set of scraping tests were carried out on MFC thin films that were pre-wetted by placing a drop (0.5 µL, 1 mm diameter) of the artificial saliva. For all scraping tests, glass substrates of 25 mm diameter were used instead of the ones of 10 mm in the contact transfer tests.

References

(1) Paton, S.; Spencer, A.; Garratt, I.; Thompson, K.-A.; Dinesh, I.; Aranega-Bou, P.; Stevenson, D.; Clark, S.; Dunning, J.; Bennett, A.; Pottage, T.; Johnson, K. N. Persistence of Severe Acute Respiratory Syndrome Coronavirus 2 (SARS-CoV-2) Virus and Viral RNA in Relation to Surface Type and Contamination Concentration. *Appl. Environ. Microbiol.* 2021, *87* (14), e00526-21.
(2) WHO *Transmission of SARS-CoV-2: implications for infection prevention precautions: scientific brief, 09 July 2020*; World Health Organization: 2020.
(3) Bean, B.; Moore, B. M.; Sterner, B.; Peterson, L. R.; Gerding, D. N.; Balfour, H. H., Jr. Survival of Influenza Viruses on Environmental Surfaces. *J. Infect. Dis.* 1982, *146* (1), 47-51.
(4) Behzadinasab, S.; Chin, A. W. H.; Hosseini, M.; Poon, L. L. M.; Ducker, W. A. SARS-CoV-2 virus transfers to skin through contact with contaminated solids. *Scientific reports* 2021, *11*, 22868.
(5) WHO. *Infection prevention and control of epidemic-and pandemic-prone acute respiratory infections in health care*, World Health Organization: 2014.
(6) Huang, Y.-J. S.; Bilyeu, A. N.; Hsu, W.-W.; Hettenbach, S. M.; Willix, J. L.; Stewart, S. C.; Higgs, S.; Vanlandingham, D. L. Treatment with dry hydrogen peroxide accelerates the decay of severe acute syndrome coronavirus-2 on non-porous hard surfaces. *Am. J. Infect. Control* 2021, *49* (10), 1252-1255.
(7) Patterson, E. I.; Prince, T.; Anderson, E. R.; Casas-Sanchez, A.; Smith, S. L.; Cansado-Utrilla, C.; Solomon, T.; Griffiths, M. J.; Acosta-Serrano, Á.; Turtle, L.; Hughes, G. L. Methods of Inactivation of SARS-CoV-2 for Downstream Biological Assays. *J. Infect. Dis.* 2020, *222* (9), 1462-1467.
(8) Anderson, E. R.; Hughes, G. L.; Patterson, E. I. Inactivation of SARS-CoV-2 on surfaces and in solution with Virusend (TX-10), a novel disinfectant. *Access Microbiol.* 2021, *3* (4).
(9) Subpiramaniyam, S. Outdoor disinfectant sprays for the prevention of COVID-19: Are they safe for the environment? *Sci. Total Environ.* 2021, *759*, 144289.
(10) Nabi, G.; Wang, Y.; Hao, Y.; Khan, S.; Wu, Y.; Li, D. Massive use of disinfectants against COVID-19 poses potential risks to urban wildlife. *Environ. Res.* 2020, *188*, 109916.
(11) Dumas, O.; Varraso, R.; Boggs, K. M.; Quinot, C.; Zock, J.-P.; Henneberger, P. K.; Speizer, F. E.; Le Moual, N.; Camargo, C. A., Jr. Association of Occupational Exposure to Disinfectants With Incidence of Chronic Obstructive Pulmonary Disease Among US Female Nurses. *JAMA Netw. Open* 2019, *2* (10), e1913563-e1913563.
(12) El-Nahhal, I.; El-Nahhal, Y. Ecological consequences of COVID-19 outbreak. *J. Water Sci. Eng.* 2020, *1*, 1-5.

(13) Koksoy Vayisoglu, S.; Oncu, E. The use of cleaning products and its relationship with the increasing health risks during the COVID-19 pandemic. *Int. J. Clin. Pract.* 2021, *75* (10), e14534.
(14) Chang, A.; Schnall, A. H.; Law, R.; Bronstein, A. C.; Marraffa, J. M.; Spiller, H. A.; Hays, H. L.; Funk, A. R.; Mercurio-Zappala, M.; Calello, D. P. Cleaning and disinfectant chemical exposures and temporal associations with COVID-19—National poison data system, United States, January 1, 2020–March 31, 2020. *Morb. Mortal. Wkly. Rep.* 2020, *69* (16), 496.
(15) Biasin, M.; Bianco, A.; Pareschi, G.; Cavalleri, A.; Cavatorta, C.; Fenizia, C.; Galli, P.; Lessio, L.; Lualdi, M.; Tombetti, E.; Ambrosi, A.; Redaelli, E. M. A.; Saulle, I.; Trabattoni, D.; Zanutta, A.; Clerici, M. UV-C irradiation is highly effective in inactivating SARS-CoV-2 replication. *Scientific reports* 2021, *11* (1), 6260.
(16) Carubelli, R.; Schneider, J. E.; Pye, Q. N.; Floyd, R. A. Cytotoxic effects of autoxidative glycation. *Free Radical Biol. Med.* 1995, *18* (2), 265-269.
(17) Warnes, S. L.; Little, Z. R.; Keevil, C. W.; Colwell, R. Human Coronavirus 229E Remains Infectious on Common Touch Surface Materials. *mBio* 2015, *6* (6), e01697-15.
(18) Abraham, J.; Dowling, K.; Florentine, S. Can Copper Products and Surfaces Reduce the Spread of Infectious Microorganisms and Hospital-Acquired Infections? *Materials* 2021, *14* (13), 3444.
(19) Hosseini, M.; Behzadinasab, S.; Benmamoun, Z.; Ducker, W. A. The viability of SARS-CoV-2 on solid surfaces. *Curr. Opin. Colloid Interface Sci.* 2021, *55*, 101481.
(20) Sustainable cleaning products market to surge to $110 billion in 2025. https://www.smithers.com/en-gb/resources/2021/feb/sustainable-cleaning-market-surge-110-billion.
(21) Poon, W. C. K.; Brown, A. T.; Direito, S. O. L.; Hodgson, D. J. M.; Le Nagard, L.; Lips, A.; MacPhee, C. E.; Marenduzzo, D.; Royer, J. R.; Silva, A. F.; Thijssen, J. H. J.; Titmuss, S. Soft matter science and the COVID-19 pandemic. *Soft Matter* 2020, *16* (36), 8310-8324.
(22) Chi, Y.; Wang, Q.; Chen, G.; Zheng, S. The Long-Term Presence of SARS-CoV-2 on Cold-Chain Food Packaging Surfaces Indicates a New COVID-19 Winter Outbreak: A Mini Review. *Front. Public Health* 2021, *9* (344).
(23) Hu, Q.; He, L.; Zhang, Y. Community Transmission via Indirect Media-To-Person Route: A Missing Link in the Rapid Spread of COVID-19. *Front. Public Health* 2021, *9* (1064).
(24) Chatterjee, S.; Murallidharan, J. S.; Agrawal, A.; Bhardwaj, R. Why coronavirus survives longer on impermeable than porous surfaces. *Phys. Fluids* 2021, *33* (2), 021701.
(25) Kasloff, S. B.; Leung, A.; Strong, J. E.; Funk, D.; Cutts, T. Stability of SARS-CoV-2 on critical personal protective equipment. *Scientific reports* 2021, *11* (1), 984.
(26) Kratzel, A.; Steiner, S.; Todt, D.; V'Kovski, P.; Brueggemann, Y.; Steinmann, J.; Steinmann, E.; Thiel, V.; Pfaender, S. Temperature-dependent surface stability of SARS-CoV-2. *J. Infect.* 2020, *81* (3), 452-482.
(27) Rakowska, P. D.; Tiddia, M.; Faruqui, N.; Bankier, C.; Pei, Y.; Pollard, A. J.; Zhang, J.; Gilmore, I. S. Antiviral surfaces and coatings and their mechanisms of action. *Commun. Mater.* 2021, *2* (1), 53.
(28) Gonçalves, M.; Kim, J. Y.; Kim, Y.; Rubab, N.; Jung, N.; Asai, T.; Hong, S.; Weon, B. M. Droplet evaporation on porous fabric materials. *Scientific reports* 2022, *12* (1), 1087, DOI: 10.1038/s41598-022-04877-w.
(29) Kim, J. Y.; Hwang, I. G.; Weon, B. M. Evaporation of inclined water droplets. *Scientific reports* 2017, *7* (1), 42848, DOI: 10.1038/srep42848.
(30) Pham, N. T.; McHale, G.; Newton, M. I.; Carroll, B. J.; Rowan, S. M. Application of the Quartz Crystal Microbalance to the Evaporation of Colloidal Suspension Droplets. *Langmuir* 2004, *20* (3), 841-847.
(31) Murray, B.; Narayanan, S. The Role of Wettability on the Response of a Quartz Crystal Microbalance Loaded with a Sessile Droplet. *Scientific reports* 2019, *9* (1), 17289.
(32) Moakes, R. J. A.; Davies, S. P.; Stamataki, Z.; Grover, L. M. Formulation of a Composite Nasal Spray Enabling Enhanced Surface Coverage and Prophylaxis of SARS-COV-2. *Advanced Materials* 2021, *33* (26), 2008304, DOI: https://doi.org/10.1002/adma.202008304.

(33) Vincent, L.; Soille, P. Watersheds in digital spaces: an efficient algorithm based on immersion simulations. *IEEE Trans. Pattern Anal. Mach. Intell.* 1991, *13* (6), 583-598.